%% file: nue-2010-v6.tex
\documentclass[aps,prl,twocolumn,showpacs,superscriptaddress,preprintnumbers]{revtex4}  
\usepackage{graphicx}  
\usepackage{dcolumn}   
\usepackage{bm}        
\usepackage{amssymb}   
\usepackage{multirow}
\usepackage{color}
\usepackage{units}

\newcommand{\etal}{{\it et al.}}
\newcommand{\pot}{\ensuremath{7.01\times10^{20}} protons-on-target}
\newcommand{\annpred}{\ensuremath{49.1\pm 7.0 {\rm (stat.)} \pm 2.7 {\rm (syst.)}}}
\newcommand{\annobs}{\ensuremath{54}}
\newcommand{\limit}{\ensuremath{0.12~(0.20)}}

\newcommand{\delmsq}[1]{\ensuremath{\Delta m^2_{ #1 }}}
\newcommand{\sinsq}[1]{\ensuremath{\sin^{2}\!\left(2\theta_{ #1 }\right)}}

\newcommand{\tonethree}{\ensuremath{\theta_{13}}}

\newcommand{\dcp}{\ensuremath{\delta_{CP}}}

\newcommand\numunue{\ensuremath{\nu_{\mu} \rightarrow \nu_{e}}}

\newcommand{\numu}{\ensuremath{\nu_{\mu}}}                   
\newcommand{\nue}{\ensuremath{\nu_{e}}}                      
\newcommand{\nutau}{\ensuremath{\nu_{\tau}}}                 
\newcommand{\anue}{\ensuremath{\overline{\nu}_{e}}}          

\newcommand {\numucc}{\mbox{\ensuremath{\nu_{\mu}}-CC}}
\newcommand {\nuecc}{\mbox{\ensuremath{\nu_{e}}-CC}}

\begin{document}


\title{New constraints on muon-neutrino to electron-neutrino transitions in MINOS}
\input nue-2010-authors.tex

\date{\today}
\preprint{FERMILAB-PUB-10-176-E}

\begin{abstract}
This letter reports results from a search for $\numu{}\rightarrow\nue{}$ transitions by the MINOS experiment based on a \unit[$7\times10^{20}$]{protons-on-target} exposure.  Our observation of \annobs{} candidate \nue{} events in the Far Detector with a background of \annpred{} events predicted by the measurements in the Near Detector requires $2\sin^{2}\!(2\theta_{13})\sin^{2}\!\theta_{23}<\limit{}$ at the 90\% C.L. for the normal (inverted) mass hierarchy at $\dcp=0$.  The experiment sets the tightest limits to date on the value of \tonethree{} for nearly all values of \dcp{} for the normal neutrino mass hierarchy and maximal $\sinsq{23}$.
\end{abstract}

\maketitle
Observations of neutrinos created in the Sun, in the Earth's atmosphere, at nuclear reactors, and by accelerators provide compelling evidence that neutrinos experience quantum mechanical mixing of their weak flavor states~\cite{ref:osc1,ref:osc3,ref:osc4,ref:osc5,ref:sol, ref:osc8,ref:minos08}.  The resulting neutrino oscillations imply that neutrinos have mass and can be represented in either mass or flavor bases, related by the $3\times 3$ PMNS neutrino mixing matrix~\cite{ref:PNMS}.  This matrix parameterizes the mixing amplitude using three angles ($\theta_{12}$, $\theta_{23}$, and $\theta_{13}$), two Majorana phases~\cite{ref:majphas}, and a phase ($\delta_{CP}$) that could give rise to charge-parity (CP) violation in the lepton sector.  The oscillation probability depends on the differences in the squared masses of the neutrino states and the ratio ($L/E$) of the distance the neutrino travels to the energy of the neutrino.  

In MINOS, the larger mass splitting dominates, and oscillations are manifested primarily as the energy dependent disappearance of muon-neutrinos.  MINOS has set the most precise measurement of the mass splitting $|\delmsq{}|= $ \unit[(2.43$\pm 0.13 )\times 10^{-3}$]{${\rm eV^{2}}$} ~\cite{ref:dm2note, ref:minos08} and requires $\sinsq{23}>0.9$ at the 90\% confidence level (C.L.).  At this mass splitting scale, it is expected that the \numu{} are changing predominantly into \nutau{}; however the sub-dominant \numunue{} transition mode is not excluded~\cite{ref:osc1}.  Such transitions would indicate a non-zero value of $\theta_{13}$, the unknown angle of the PMNS matrix and could open the possibility of observing CP violation in the lepton sector.  In this letter, we report new results from the search for $\numu{}\rightarrow\nue{}$ transitions.

The most stringent constraint on $\theta_{13}$, from the CHOOZ reactor experiment~\cite{ref:chooz}, implies $\sinsq{13}<0.15$ at the 90\% C.L. for the value of $|\delmsq{}|$ measured by MINOS.  However, a recent global analysis of oscillation measurements hints at a non-zero value for $\theta_{13}$~\cite{ref:fogli}.  The CHOOZ limit is based on a measurement of the probability for electron-antineutrino disappearance. MINOS measures the probability of electron-neutrino appearance, which additionally depends on  $\sin^{2}\!\theta_{23}$, $\delta_{CP}$, and the sign of $\delmsq{}$.  MINOS is the first experiment to probe $\sinsq{13}$ with sensitivity beyond the CHOOZ limit.  An initial measurement with \unit[$3.14\times 10^{20}$]{protons-on-target (POT)} yielded 35 observed \nue{}-like events with an expected background of $27\pm5{\rm (stat.)}\pm2{\rm (syst.)}$ events~\cite{ref:nueprl}.  This $1.5\sigma$ excess of events is consistent with a value of \sinsq{13} near the CHOOZ limit.  The present analysis is based on an integrated exposure of \pot{} and includes the data set from the previous analysis.  

In MINOS, interactions of neutrinos produced in the Fermilab NuMI beam line~\cite{ref:beam} are observed in two detectors: a Near Detector (ND) with a \unit[29]{t} fiducial mass \unit[1.04]{km} from the production target and a Far Detector (FD) with a \unit[4]{kt} fiducial mass \unit[735]{km} from the target.  Both detectors are magnetized tracking calorimeters, composed of planes of \unit[2.54]{cm} thick steel and \unit[1.0]{cm} thick scintillator ($1.4$ radiation lengths per plane). The scintillator planes are segmented into \unit[4.1]{cm} wide strips ($1.1$ Moli\`ere radii)~\cite{ref:minosnim}.  The high statistics data set collected at the ND establishes the properties of the mostly \numu{} beam before oscillations.  The signature of $\numu{}\rightarrow\nue{}$  oscillations is an excess of \nue{} interactions in the FD relative to the expected background based on the ND observation.   

Neutrino flavor can be identified in charged current (CC) interactions by the event topology produced by the associated charged lepton.  Muons deposit energy consistent with a minimum ionizing particle that can be tracked through successive detector planes (a track).  Electrons, on the other hand, deposit energy in a relatively narrow and short region (an electromagnetic shower).  Additional detector activity can be produced by the breakup of the recoil nucleus and other particles produced in the interaction.  

In this analysis, the dominant backgrounds to \nuecc{} events are neutral current (NC) interactions and \numucc{} interactions with low energy muons.  These interactions can produce signatures that are similar to those of \nuecc{} events, especially when the hadronic system includes a $\pi^{0}$.  An irreducible background arises from the 1.3\% \nue{}+\anue{} component of the beam.  This beam \nue{} background results primarily from decays of muons produced in pion and kaon decays. Their rate below \unit[8]{GeV} is well constrained by the measured \numu{} energy spectrum~\cite{ref:minos08, ref:PRD}.  Smaller background components come from cosmogenic sources and CC interactions of \nutau{} coming from $\numu{}\rightarrow\nutau{}$ oscillations.

Selection criteria are applied to events to isolate \nuecc{} interactions and suppress backgrounds.  Cosmogenic backgrounds in this analysis are reduced to less than 0.3 events (90\% C.L.) in the FD by applying directional requirements and requiring the events to be in time with the accelerator pulse.  Selected events must have reconstructed energy between 1 and \unit[8]{GeV}, a reconstructed shower, and at least 5 contiguous planes, each with energy depositions above half the energy deposited by a minimum ionizing particle.  Events with long tracks are rejected. Further enrichment is achieved using an artificial neural network (ANN) with 11 input variables characterizing the longitudinal and transverse energy deposition in the calorimeter~\cite{ref:tingjun}.  The variables used in the ANN are identical to those used in~\cite{ref:nueprl}, but the network was re-optimized over a sample of simulated events generated with a refined detector response model, improved event reconstruction, and better modeling of hadron scattering within the iron nucleus.  Maximum sensitivity is achieved by selecting events with the neural network output above $0.7$.  Background rejection is improved by a factor of 1.2 over that reported in~\cite{ref:nueprl} for a similar signal efficiency.  

The number of expected background events is determined from ND data.  The extrapolation to the FD of each of the primary background components, \numucc{}, NC, and beam \nuecc{}, has a different dependence on oscillation probability and beam geometry and is treated separately.  Individual background components are determined using three beam configurations, each with different relative background compositions.  The first configuration is the standard one used for the appearance search.  The hadron production target is located close to the first focusing horn, producing a neutrino beam peaked at \unit[3]{GeV}.  In the second configuration, the target is moved upstream from the horns causing higher energy hadrons to be focused and yielding a neutrino spectrum peaked at \unit[9]{GeV}.  In a third configuration the current in the focusing horns is turned off so no hadrons are focused.  Consequently, the low-energy peak of the neutrino energy distribution disappears, and the selected event sample is dominated by NC events from higher energy neutrino interactions.  

Data obtained in the above configurations and the simulated ratios of rates for each configuration are used to extract the three individual background spectra.  The beam line, detector, and particle propagation simulation is based on {\tt GEANT3}~\cite{ref:geant} and the hadron production yields from the target are based on {\tt FLUKA}~\cite{ref:fluka}.  Neutrino interactions and further re-interactions of the resulting hadrons within the nucleus are simulated using {\tt NEUGEN3}~\cite{ref:neugen}.  The predicted neutrino energy spectrum is adjusted to agree with the ND \numucc{} data~\cite{ref:PRD}. 

\begin{figure}
\begin{center}
\includegraphics[viewport=10 50  567 427, keepaspectratio,width= 0.48 \textwidth, clip=true]{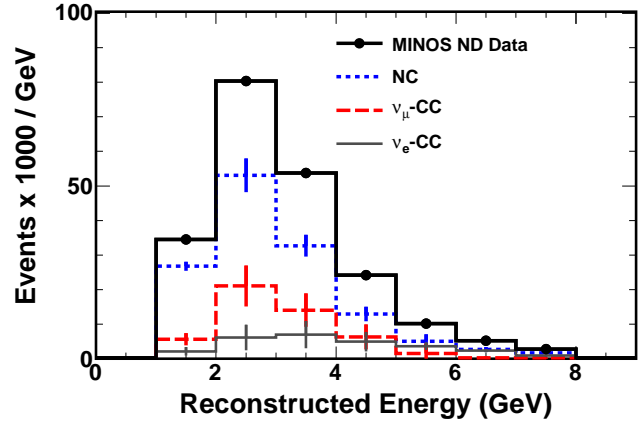}
\end{center}
\caption{Reconstructed Near Detector energy spectra of the \nuecc{} selected events (bold solid).  Also shown is the decomposition of this spectrum into neutral current (dotted), \numu{} charged current (dashed), and beam \nuecc{} (light solid) components determined using the multiple beam configuration method.   The sum of the three background components is constrained to agree with the data.  Uncertainties on the data are statistical and are not visible on this scale; uncertainties on the components are systematic.}
\label{fig:nddata}
\end{figure}

Data were collected during three run periods, each with somewhat different beam conditions.  Most notably, during the third run period, the decay pipe was filled with helium at \unit[0.9]{atm} for safety reasons.  The background decomposition is performed as a function of neutrino energy and is done separately for each run period to account for the different beam conditions, a small, gradual target degradation, and detector aging.  Figure~\ref{fig:nddata} shows the energy spectrum measured in the ND for events passing the selection criteria and the extracted NC,  \numucc{}, and beam \nuecc{} components.  The ND background is (64$\pm$5)\% NC, (23$\pm$5)\% \numucc{} and (13$\pm$3)\% beam \nuecc{} events.  The errors on the components are derived primarily from the data and are correlated due to the constraint that the background must add up to the observed ND event rate.  This constraint also leads to a much reduced error on the FD prediction.   A second decomposition technique was applied to verify the background components.   This method uses \numucc{} events with the muon track removed.  The remnant hits are then processed through the standard analysis~\cite{ref:nueprl,ref:AHthesis}.  This second method yields consistent ND background components.

After decomposition of the ND data from each run period into separate background components, each spectrum is multiplied by the ratio of FD to ND event rates in reconstructed energy bins based on the simulation for that component, providing a prediction of the FD spectrum in the absence of \nue{} appearance.  Neutrino oscillations are included when predicting the FD event rate.  The predictions are summed to give 49.1 expected background events, of which 35.8 are NC, 6.3 \numucc{}, 5.0 beam \nue{} and 2.0 \nutau{}~\cite{ref:oscpars}.

The efficiency for selecting \nuecc{} events is estimated using remnants from muon-removed \numucc{} events with a simulated electron replacing the muon.  The embedded electron has the same momentum and direction as the removed muon.  Test beam measurements~\cite{ref:caldet} demonstrate that single electrons are well modeled in the MINOS detectors, and the selection efficiency of electrons agrees with the simulation to within 1.6\%.  With this method, we find our efficiency for selecting \nuecc{} events to be (41.6$\pm$1.0)\%.

\begin{figure}
\begin{center}
\includegraphics[viewport=50 50  567 411, keepaspectratio,width= 0.45 \textwidth, clip=true]{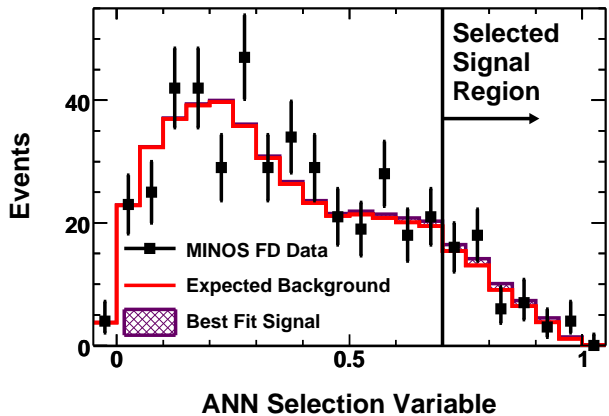}
\end{center}
\caption{Distribution of the ANN selection variable for pre-selected events in the Far Detector.  Black points show data with statistical error bars. The non-shaded histogram shows the expected background. The shaded region shows the additional \nue{} charged current events allowed from the best fit to the oscillation hypothesis as described in the text.}
\label{fig:fdpid}
\end{figure}

Systematic uncertainties are evaluated by generating Monte Carlo (MC) samples in which systematic effects are varied over their expected range of uncertainty and quantifying the change in the number of predicted background events in the FD.  Most of the dominant uncertainties arise from Far/Near differences.  The principal systematic effects, listed in Table~\ref{tab:systematics}, include (a) uncertainties in energy scale, (b) uncertainty in Near to Far relative event rate, and (c) uncertainties in the nuclear hadronization and intranuclear scattering models.  Other systematic error sources (d) including uncertainties in cross section models, beam flux, and the details of the detector simulation each contribute to the systematic error at lower levels.  While uncertainties in the composition and kinematic distribution of the particles that emerge from the nucleus can be large, these and other uncertainties associated with neutrino interaction physics mostly cancel when comparing the ND and FD data.  The use of the same materials and detector segmentation in the ND and FD is critical in achieving this error cancellation.  The individual systematic errors on the expected background are combined in quadrature with the uncertainty from the decomposition of the background and a systematic error on the \nutau{} background to give an overall systematic uncertainty of 5.6\% on the expected number of background events in the FD.  

\begin{table}
\begin{tabular}{l l c}
\hline 
\multirow{2}{*}{Uncertainty source} & & Uncertainty on\\
 & &  background events \\
\hline 
Far/Near ratio: &&4.5\%\\
\ \ (a) Energy Scale &2.8\%&\\
\ \ (b) Relative Event Rate  & 2.4\%&\\
\ \ (c) Hadronic Model &2.5\%&\\
\ \ (d) All Other Combined & 0.7\%&\\
\hline
Near Detector Decomposition & &2.8\% \\
\nutau{} background & &1.7\%\\
\hline

Total Systematic Uncertainty && 5.6\% \\
Expected Statistical Uncertainty && 14.3\% \\
\hline
\end{tabular}
\caption{Systematic uncertainty in the total number of background events in the Far Detector.}
\label{tab:systematics}
\end{table}

The expected number of background events and its uncertainty, along with all the analysis procedures are established before examining the full FD data set.  Additionally, before counting events in the signal region of ANN $>$ 0.7, two FD data samples are examined to check the expected event rate and the background rejection in the FD.   To verify the expected event rate, the full decomposition and extrapolation method is applied to events well below the signal region (${\rm ANN}<0.5$), giving a prediction of 313.6 events.  We observe 327 events, consistent with the prediction to within $1\sigma$.  Background rejection is verified by examining muon-removed \numucc{} events.  In the FD ($92.8\pm0.9 ~{\rm(stat.)}$)\% of muon removed events are rejected, in agreement with ($93.58\pm0.05 ~{\rm(stat.)}$)\% predicted from the ND data.

\begin{figure}
\begin{center}
\includegraphics[viewport=50 50  567 411, keepaspectratio,width= 0.45 \textwidth, clip=true]{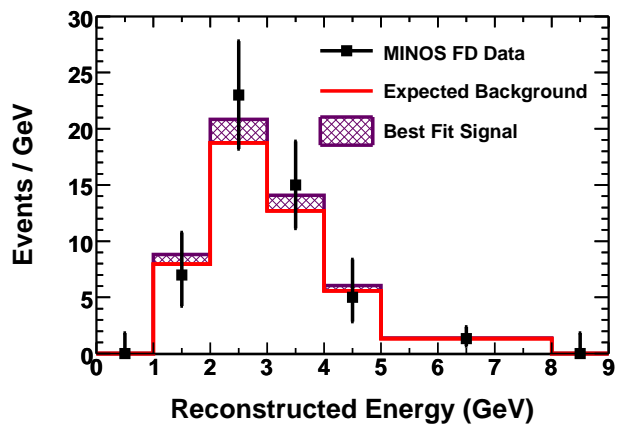}
\end{center}
\caption{Reconstructed energy spectrum of the events in the Far Detector which pass all \nue{} charged current selection criteria, with the exception of the energy cut. Black points show data with statistical error bars.  The non-shaded histogram shows the expected background. The shaded region shows the additional \nue{} charged current events allowed by the best fit to the oscillation hypothesis.}
\label{fig:fdspectrum}
\end{figure}

Figure~\ref{fig:fdpid} shows the number of selected candidate events in the FD as a function of the ANN selection variable.  The energy spectrum for the events in the signal region is shown in Figure~\ref{fig:fdspectrum}.  We observe \annobs{} events in the signal region with an expected background of \annpred{}, a $0.7\sigma$ excess over the expected background.  Similar results were produced by a cross check analysis that used a different neural network based on an alternate event reconstruction algorithm~\cite{ref:stevethesis}. Taking into account the improved background rejection in the current analysis, this result is consistent with the earlier report, based on a smaller data sample. From that sample we now select 28 events with an expected background of $22.5\pm4.7{\rm (stat.)}\pm1.1{\rm (syst.)}$ events.

\begin{figure}
\begin{center}
\includegraphics[viewport=40 40 464 680, keepaspectratio,width= 0.45 \textwidth, clip=true]{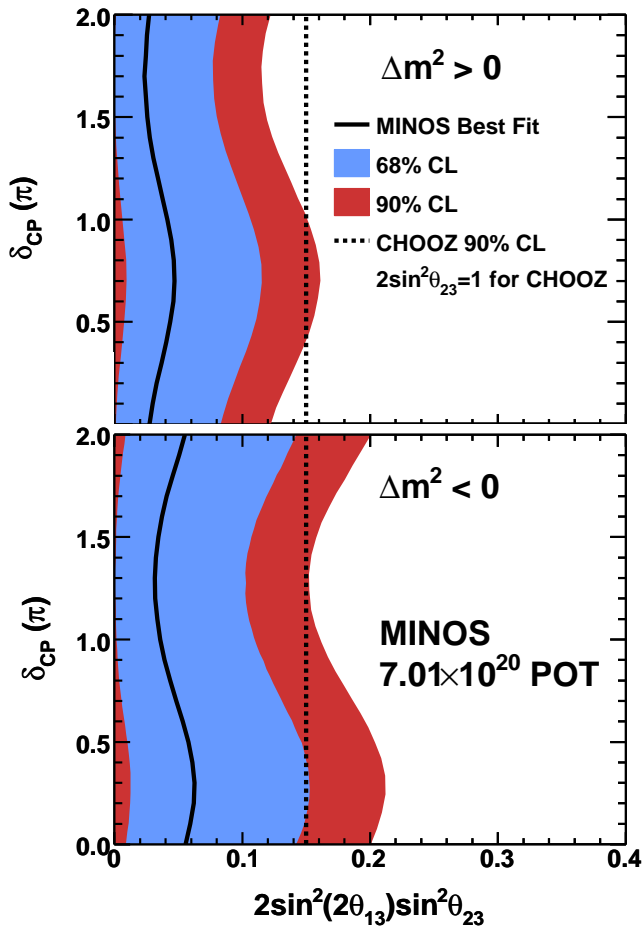}
\end{center}
\caption{Values of $2\sinsq{13}\sin^{2}\!\theta_{23}$ and $\delta_{CP}$ that produce a number of candidate events in the Far Detector consistent with the observation for the normal hierarchy (top) and inverted hierarchy (bottom).  Black lines show those values that best represent our data. Red (blue) regions show the 90\% (68\%) C.L. intervals.  The CHOOZ limit is drawn for \delmsq{32}=$2.43\times10^{-3} {\rm eV^{2}}$, \sinsq{23}=1.0}
\label{fig:sens}
\end{figure}

Figure~\ref{fig:sens} shows the values of $2\sinsq{13}\sin^{2}\!\theta_{23}$ and $\delta_{CP}$ that give a number of events consistent with our observation.  The oscillation probability is computed using a full 3-flavor neutrino mixing framework with matter effects~\cite{ref:3flav,ref:JBthesis}, which includes a dependence on the neutrino mass hierarchy.  Statistical and systematic uncertainties are included when constructing the confidence intervals via the Feldman-Cousins approach~\cite{ref:FC}.  The variations of the values of  $|\delmsq{32}|$, \delmsq{21}, $\sin^{2}\!\theta_{23}$, and \sinsq{12} within their experimental errors~\cite{ref:osc1, ref:osc8, ref:minos08, ref:sol} are included in the computation of the contours.

In conclusion, we report improved constraints on $2\sin^{2}\!(2\theta_{13})\sin^{2}\!\theta_{23}$ from the search for \nue{} appearance by the MINOS experiment. The \annobs{} events selected in the Far Detector are 0.7\,$\sigma$ higher than the expected background of \annpred{}.  Interpreted as an upper limit on the probability of \numunue{} oscillations, our data require $2\sin^{2}\!(2\theta_{13})\sin^{2}\!\theta_{23}<\limit{}$ at the 90\% C.L. at $\dcp{}=0$ for the normal (inverted) hierarchy.  This measurement represents the best constraint on the value of \tonethree{} for nearly all values of \dcp{} assuming the normal mass hierarchy and maximal \sinsq{23}.  

This work was supported by the US DOE; the UK STFC; the US NSF; the State and University of Minnesota; the University of Athens, Greece; and Brazil's FAPESP, CNPq, and CAPES.  We are grateful to the Minnesota DNR, the crew of the Soudan Underground Laboratory, and the staff of Fermilab for their contributions to this effort.

\end{document}

%% file: nue-2010-authors.tex
\newcommand{\Cambridge}{Cavendish Laboratory, University of Cambridge, Madingley Road, Cambridge CB3 0HE, United Kingdom}
\newcommand{\FNAL}{Fermi National Accelerator Laboratory, Batavia, Illinois 60510, USA}
\newcommand{\RAL}{Rutherford Appleton Laboratory, Science and Technology Facilities Council, OX11 0QX, United Kingdom}
\newcommand{\UCL}{Department of Physics and Astronomy, University College London, Gower Street, London WC1E 6BT, United Kingdom}
\newcommand{\Caltech}{Lauritsen Laboratory, California Institute of Technology, Pasadena, California 91125, USA}
\newcommand{\ANL}{Argonne National Laboratory, Argonne, Illinois 60439, USA}
\newcommand{\Athens}{Department of Physics, University of Athens, GR-15771 Athens, Greece}
\newcommand{\NTUAthens}{Department of Physics, National Tech. University of Athens, GR-15780 Athens, Greece}
\newcommand{\Benedictine}{Physics Department, Benedictine University, Lisle, Illinois 60532, USA}
\newcommand{\BNL}{Brookhaven National Laboratory, Upton, New York 11973, USA}
\newcommand{\CdF}{APC -- Universit\'{e} Paris 7 Denis Diderot, 10, rue Alice Domon et L\'{e}onie Duquet, F-75205 Paris Cedex 13, France}
\newcommand{\Cleveland}{Cleveland Clinic, Cleveland, Ohio 44195, USA}
\newcommand{\Delhi}{Department of Physics and Astrophysics, University of Delhi, Delhi 110007, India}
\newcommand{\GEHealth}{GE Healthcare, Florence South Carolina 29501, USA}
\newcommand{\Harvard}{Department of Physics, Harvard University, Cambridge, Massachusetts 02138, USA}
\newcommand{\HolyCross}{Holy Cross College, Notre Dame, Indiana 46556, USA}
\newcommand{\IIT}{Physics Division, Illinois Institute of Technology, Chicago, Illinois 60616, USA}
\newcommand{\Iowa}{Department of Physics and Astronomy, Iowa State University, Ames, Iowa 50011, USA}
\newcommand{\Indiana}{Indiana University, Bloomington, Indiana 47405, USA}
\newcommand{\ITEP}{High Energy Experimental Physics Department, ITEP, B. Cheremushkinskaya, 25, 117218 Moscow, Russia}
\newcommand{\JMU}{Physics Department, James Madison University, Harrisonburg, Virginia 22807, USA}
\newcommand{\LASL}{Nuclear Nonproliferation Division, Threat Reduction Directorate, Los Alamos National Laboratory, Los Alamos, New Mexico 87545, USA}
\newcommand{\Lebedev}{Nuclear Physics Department, Lebedev Physical Institute, Leninsky Prospect 53, 119991 Moscow, Russia}
\newcommand{\LLL}{Lawrence Livermore National Laboratory, Livermore, California 94550, USA}
\newcommand{\MIT}{Lincoln Laboratory, Massachusetts Institute of Technology, Lexington, Massachusetts 02420, USA}
\newcommand{\Minnesota}{University of Minnesota, Minneapolis, Minnesota 55455, USA}
\newcommand{\Crookston}{Math, Science and Technology Department, University of Minnesota -- Crookston, Crookston, Minnesota 56716, USA}
\newcommand{\Duluth}{Department of Physics, University of Minnesota -- Duluth, Duluth, Minnesota 55812, USA}
\newcommand{\Otterbein}{Otterbein College, Westerville, Ohio 43081, USA}
\newcommand{\Oxford}{Subdepartment of Particle Physics, University of Oxford, Oxford OX1 3RH, United Kingdom}
\newcommand{\Pittsburgh}{Department of Physics and Astronomy, University of Pittsburgh, Pittsburgh, Pennsylvania 15260, USA}
\newcommand{\IHEP}{Institute for High Energy Physics, Protvino, Moscow Region RU-140284, Russia}
\newcommand{\RoyalH}{Physics Department, Royal Holloway, University of London, Egham, Surrey, TW20 0EX, United Kingdom}
\newcommand{\Carolina}{Department of Physics and Astronomy, University of South Carolina, Columbia, South Carolina 29208, USA}
\newcommand{\SLAC}{Stanford Linear Accelerator Center, Stanford, California 94309, USA}
\newcommand{\Stanford}{Department of Physics, Stanford University, Stanford, California 94305, USA}
\newcommand{\StJohnFisher}{Physics Department, St. John Fisher College, Rochester, New York 14618 USA}
\newcommand{\Sussex}{Department of Physics and Astronomy, University of Sussex, Falmer, Brighton BN1 9QH, United Kingdom}
\newcommand{\TexasAM}{Physics Department, Texas A\&M University, College Station, Texas 77843, USA}
\newcommand{\Texas}{Department of Physics, University of Texas at Austin, 1 University Station C1600, Austin, Texas 78712, USA}
\newcommand{\TechX}{Tech-X Corporation, Boulder, Colorado 80303, USA}
\newcommand{\Tufts}{Physics Department, Tufts University, Medford, Massachusetts 02155, USA}
\newcommand{\UNICAMP}{Universidade Estadual de Campinas, IFGW-UNICAMP, CP 6165, 13083-970, Campinas, SP, Brazil}
\newcommand{\USP}{Instituto de F\'{i}sica, Universidade de S\~{a}o Paulo,  CP 66318, 05315-970, S\~{a}o Paulo, SP, Brazil}
\newcommand{\Warsaw}{Department of Physics, University of Warsaw, Ho\.{z}a 69, PL-00-681 Warsaw, Poland}
\newcommand{\Washington}{Physics Department, Western Washington University, Bellingham, Washington 98225, USA}
\newcommand{\WandM}{Department of Physics, College of William \& Mary, Williamsburg, Virginia 23187, USA}
\newcommand{\Wisconsin}{Physics Department, University of Wisconsin, Madison, Wisconsin 53706, USA}
\newcommand{\deceased}{Deceased.}

\affiliation{\ANL}
\affiliation{\Athens}
\affiliation{\Benedictine}
\affiliation{\BNL}
\affiliation{\Caltech}
\affiliation{\Cambridge}
\affiliation{\UNICAMP}
\affiliation{\FNAL}
\affiliation{\Harvard}
\affiliation{\HolyCross}
\affiliation{\IIT}
\affiliation{\Indiana}
\affiliation{\Iowa}
\affiliation{\Lebedev}
\affiliation{\LLL}
\affiliation{\UCL}
\affiliation{\Minnesota}
\affiliation{\Duluth}
\affiliation{\Otterbein}
\affiliation{\Oxford}
\affiliation{\Pittsburgh}
\affiliation{\RAL}
\affiliation{\USP}
\affiliation{\Carolina}
\affiliation{\Stanford}
\affiliation{\Sussex}
\affiliation{\TexasAM}
\affiliation{\Texas}
\affiliation{\Tufts}
\affiliation{\Warsaw}
\affiliation{\WandM}

\author{P.~Adamson}
\affiliation{\FNAL}

\author{C.~Andreopoulos}
\affiliation{\RAL}



\author{D.~J.~Auty}
\affiliation{\Sussex}


\author{D.~S.~Ayres}
\affiliation{\ANL}

\author{C.~Backhouse}
\affiliation{\Oxford}




\author{G.~Barr}
\affiliation{\Oxford}





\author{R.~H.~Bernstein}
\affiliation{\FNAL}

\author{M.~Betancourt}
\affiliation{\Minnesota}


\author{P.~Bhattarai}
\affiliation{\Duluth}

\author{M.~Bishai}
\affiliation{\BNL}

% ...

\collaboration{The MINOS Collaboration}
\noaffiliation

%% file: nue-2010-v6.bbl
\begin{thebibliography}{99}

\bibitem{ref:osc1}Y. Ashie \etal{}, Phys. Rev. Lett. {\bf 93}, 101801 (2004); Phys. Rev. D {\bf 71}, 112005 (2005).
\bibitem{ref:osc3}W.W.M. Allison \etal{}, Phys. Rev. D {\bf 72}, 052005 (2005).
\bibitem{ref:osc4}M. Ambrosio \etal{}, Eur. Phys. J. C {\bf 36}, 323 (2004).
\bibitem{ref:osc5}M.H. Ahn \etal{}, Phys Rev D {\bf 74}, 072003 (2006).
\bibitem{ref:sol}{B. Aharmim \etal{}, Phys. Rev. C {\bf 72} 055502 (2005).}
\bibitem{ref:osc8}T. Araki \etal{}, Phys. Rev. Lett. {\bf 94}, 081801 (2005).
\bibitem{ref:minos08}{P. Adamson \etal{}, Phys. Rev. Lett. {\bf 101}, 131802 (2008).}

\bibitem{ref:PNMS}
{B. Pontecorvo, JETP {\bf 34}, 172 (1958); V.N. Gribov and B. Pontecorvo, Phys. Lett. B {\bf 28}, 493 (1969); Z. Maki, M. Nakagawa, and S. Sakata, Prog. Theor. Phys. {\bf 28}, 870 (1962).}
\bibitem{ref:majphas} Oscillation experiments are not sensitive to the Majorana phases.
\bibitem{ref:dm2note}
{The experiment measures an unresolved mixture of $|\delmsq{31}|$ and $|\delmsq{32}|$ which we refer to as $|\delmsq{}|$ for brevity.  For further discussion see G. Fogli \etal{}, Prog. Part. Nucl. Phys. {\bf 57}, 742 (2006).}
\bibitem{ref:chooz}
{M. Apollonio \etal{}, Eur. Phys. J. C {\bf 27}, 331 (2003).}
\bibitem{ref:fogli}
{G.L. Fogli \etal{},  Phys. Rev. Lett.\ {\bf 101}, 141801 (2008).}
\bibitem{ref:nueprl}
{P. Adamson \etal{}, Phys. Rev. Lett. {\bf 103}, 261802 (2009).}
\bibitem{ref:beam}
{S. Kopp, Proc. 2005 IEEEPart. Accel. Conf., May 2005, Fermilab-Conf-05-093-AD and arXiv:physics/0508001.}
\bibitem{ref:minosnim}
{D.G. Michael \etal{},  Nucl. Inst. \& Meth. A {\bf 596}, 190 (2008). }
\bibitem{ref:PRD} {P. Adamson \etal{}, Phys. Rev. D {\bf 77}, 072002 (2008). }
\bibitem{ref:tingjun}{T. Yang, Ph.D. Thesis, Stanford University (2009).}
\bibitem{ref:geant}
{R. Brun \etal{}, CERN Program Library W5013 (1984).}
\bibitem{ref:fluka}
{A. Fasso \etal{}, CERN-2005-10 (2005).}
\bibitem{ref:neugen} 
{S. Dytman, H. Gallagher, and M. Kordosky, arXiv:0806.2119 (2008). }
\bibitem{ref:AHthesis} {A. Holin, Ph.D. Thesis, UCL (2010).}
\bibitem{ref:oscpars}
{For \delmsq{32}=$2.43\times10^{-3} {\rm eV^{2}}$, \sinsq{23}=1.0, and \sinsq{13}=0.} 
\bibitem{ref:caldet}
{P. Adamson \etal{}, Nucl. Inst. \& Meth. A {\bf 556}, 119 (2006); A. Cabrera \etal{}, Nucl. Inst. \& Meth. A {\bf 609}, 106 (2009).}
\bibitem{ref:stevethesis}
{S. Cavanaugh, Ph.D. Thesis, Harvard University (2010).}
\bibitem{ref:3flav}
{E.K. Akhmedov \etal{}, J. High En. Phys. {\bf 0404}, 078 (2004).}
\bibitem{ref:JBthesis}{J. Boehm, Ph.D. Thesis, Harvard University (2009).}
\bibitem{ref:FC}
{G.J. Feldman and R.D. Cousins, Phys. Rev. D {\bf 57}, 3873 (1998).}
\end{thebibliography}
